\documentclass{sobolev}    
\setpg{1}                  

\pdfoutput=1

\titl[Magnetic Fields in Massive Stars]{Magnetic Fields in Massive Stars: \\ New Insights} 

\authors{S.~Hubrig, M.~Sch\"oller et al.}{S.~Hubrig\aff{1}, M.~Sch\"oller\aff{2}, A.F.~Kholtygin\aff{3}, 
L.M.~Oskinova\aff{4}, I.~Ilyin\aff{1}}

\affiliat{
 \item  Leibniz-Institut f\"ur Astrophysik Potsdam (AIP), An der Sternwarte~16, 14482 Potsdam, Germany
\item European Southern Observatory, Karl-Schwarzschild-Str.~2, 85748 Garching, Germany
 \item Astronomical Institute, Saint-Petersburg State University, Saint-Petersburg, Russia
 \item Institut f\"ur Physik und Astronomie, Universit\"at Potsdam, Karl-Liebknecht-Str.~24/25, 14476 Potsdam, Germany
 }

\email{shubrig@aip.de}   

\begin{document} 

\begin{abstract}
Substantial progress has been achieved over the last decade in studies of stellar magnetism 
due to the improvement of magnetic field measurement methods.
We review recent results on the magnetic field characteristics of early B- and O-type
stars obtained by various teams using different measurement techniques. 
\end{abstract}  

\section{Massive O-type stars with different spectral designations and kinematic characteristics}\label{otype}

During the last years, a number of magnetic studies focused on the detection of magnetic fields
in massive early B and O-type stars.
The characterization of magnetic fields in massive stars is indispensable to understand the conditions controlling 
the presence of those fields and their implications for the stellar physical parameters and evolution. 
Accurate studies of the age, environment, and 
kinematic characteristics of magnetic stars are also promising to give us new insights into the origin 
of the magnetic fields.
While a number of early B-type stars were detected as magnetic already several decades back,
the first magnetic field detection in an O-type star was achieved only 13 years ago,
even though the existence of magnetic O-type stars had been suspected for a long time.
Indirect observational evidence for the presence of magnetic fields were the many unexplained phenomena 
observed in massive stars, which are thought to be related to magnetic fields,
like cyclical wind variability, H$\alpha $ emission variation, 
chemical peculiarity, narrow X-ray emission lines, and non-thermal radio/X-ray emission.

However, direct measurements of the magnetic field strength in massive stars, using spectropolarimetry 
to determine the Zeeman splitting of the spectral lines, are difficult, since only a few spectral lines are available 
for these measurements.
In addition, these spectral lines are usually strongly broadened by rapid rotation and macroturbulence,
and frequently appear in emission or display P\,Cyg profiles.
In high-resolution spectropolarimetric observations, broad spectral lines frequently extend 
over adjacent orders, so that it is necessary to adopt order shapes to get the best continuum normalization.
Furthermore, most of the existing high-resolution spectropolarimeters are operating at smaller telescopes, and cannot
deliver the necessary high signal-to-noise (SNR) observations for a majority of the massive stars.
Especially O-type stars and Wolf-Rayet (WR) stars are rather faint. 
Indeed, the Bright Star Catalog contains only about 50 O-type stars and only very few WR stars.

In view of the large line 
broadening in massive stars, to search for the presence of magnetic fields, the low-resolution 
VLT instrument FORS\,2 --
and prior to that FORS\,1 -- appears to be the most suitable instrument in the world,
offering the appropriate spectral resolution and the required spectropolarimetric sensitivity,
giving access to massive stars even in galaxies in our neighborhood.
Only the Faint Object Camera and Spectrograph at the Subaru Telescope 
has an operating spectropolarimetric mode, and,
pending the commissioning of the PEPSI spectrograph in polarimetric mode installed at the Large Binocular 
Telescope, no further high-resolution spectropolarimetric capabilities are available on any of the 8--10 m class 
telescopes.

\begin{figure}
\centering
\includegraphics[width=0.45\textwidth]{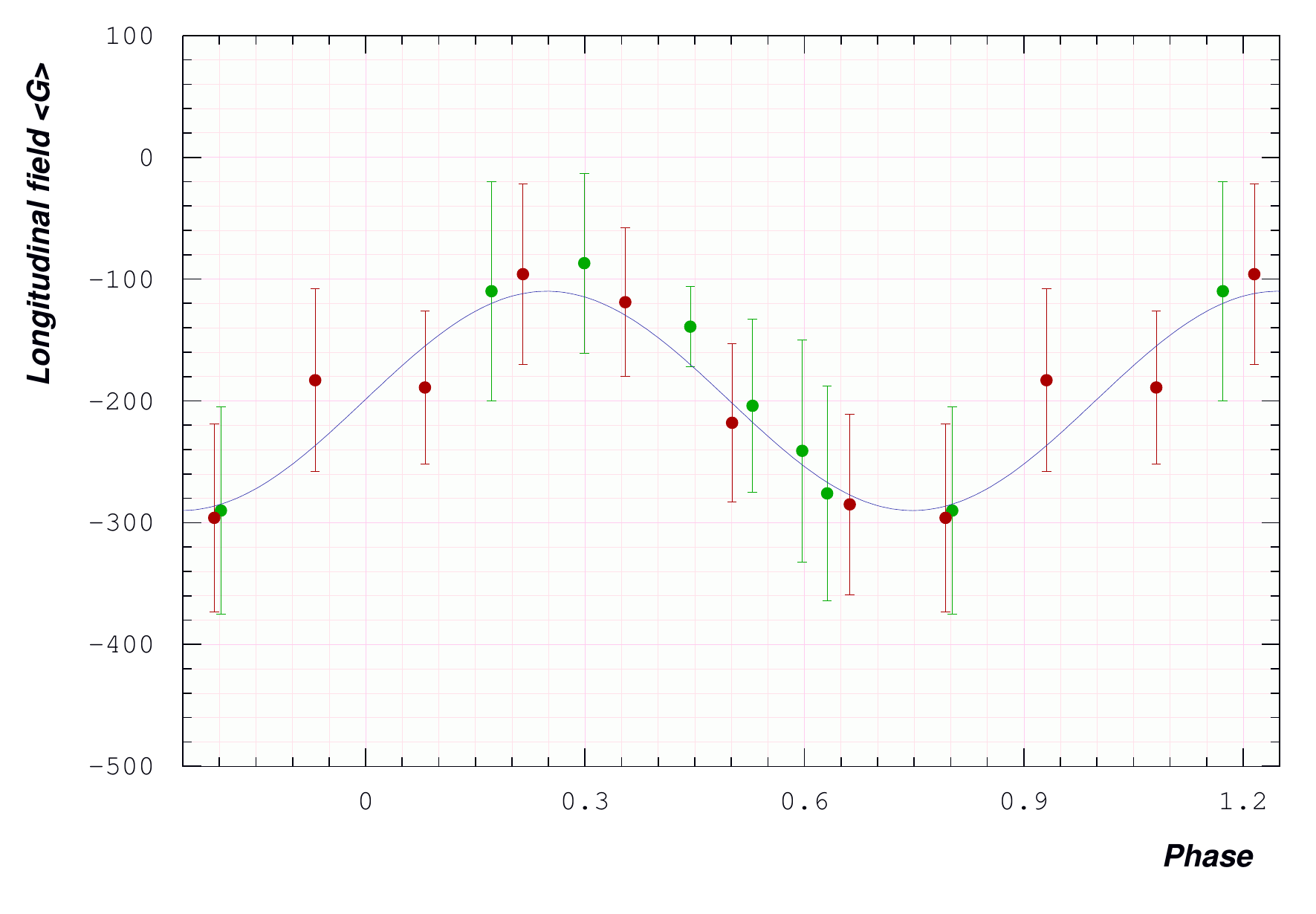}
\caption{
 Longitudinal magnetic field variation of the Of?p star HD\,148937 according to the 7.032\,d period 
determined by Naz\'e et al.\ \cite{Naze}.
Red symbols correspond to ESPaDOnS observations \cite{Wade},
while green symbols are FORS\,1 and FORS\,2 measurements \cite{Hubrig2008,Hubrig2013}.
Note that the measurement errors for both ESPaDOnS and FORS\,1/2 observations are of similar order.
}
\label{fig:ofp}
\end{figure}

The first spectropolarimetric observations of O-type stars at ESO started with FORS\,1 already in 2005.
During a survey of thirteen O-type stars, the discovery of the presence of a magnetic field was announced
in the Of?p star HD\,148937 \cite{Hubrig2008}.
The class of Of?p stars was introduced by Walborn \cite{Walborn} and includes only five stars in our Galaxy.
Of?p stars display recurrent spectral variations in 
certain spectral lines, sharp emission or P\,Cygni profiles in He\,{\sc i} and the Balmer lines, and strong C\,{\sc iii}
emission lines around 4650\,\AA{}.
In the last years, it was shown that all Of?p stars are magnetic with field strengths
from a few hundred Gauss to a few kG.
Among them, only two Of?p stars, HD\,148937 and CPD$-$28\,2561 are observable from Paranal and, noteworthy, the 
first magnetic field detections were achieved through FORS\,1 and FORS\,2 observations \cite{Hubrig2013}.

All FORS\,1/2 observations of HD\,148937 are presented in Fig.~\ref{fig:ofp} together with the ESPaDOnS 
observations obtained at CFHT \cite{Wade}.
This figure demonstrates the excellent agreement between the FORS\,2 and  ESPaDOnS measurements,
highlighting the outstanding potential of FORS\,2 for the
detection of magnetic fields and the investigation of the magnetic field geometry in massive stars. 
Notably, while an exposure time of 21.5\,h at the CFHT was necessary to obtain seven binned measurements, the exposure 
time for the individual FORS\,2 observations accounted only for two to four 
minutes and only 2.3\,h  were used 
for the observations at six different epochs, including telescope presets and the usual overheads for readout time and 
retarder waveplate rotation. 

Also the FORS\,2 measurements
of the mean longitudinal magnetic field of the second Of?p star, CPD$-$28\,2561, 
were consistent with a single-wave variation during the stellar rotation cycle,
indicating a dominant dipolar contribution to the magnetic field topology with an estimated polar
strength of the surface dipole B$_d$ larger than 1.15 kG \cite{Hubrig2015a}. 
Interestingly, in the studies of these two Of?p stars,
none of the reported detections reached a 4$\sigma$ significance level.
While 3$\sigma$ detections with FORS\,2 can not always be trusted 
for single observations, they are genuine if the measurements show
smooth variations over the rotation period, similar to those found for the Of?p
stars HD\,148937 and CPD$-$28\,2561.
The detection of rotational modulation of the 
longitudinal magnetic field is important to constrain the global field geometry
necessary to support physical modeling of the spectroscopic and light variations.

To identify and to model the physical processes responsible for the generation of their magnetic fields,
it is important to establish whether magnetic fields can also be detected in massive stars that are fast rotators
and have runaway status. Recent detections of strong magnetic fields in very fast rotating early-B 
type stars indicate that the spindown timescale via magnetic braking can be much longer than the estimated 
age of these targets (e.g.\ \cite{rivinius}).
Furthermore, current studies of their kinematical status identified a number of magnetic O and Of?p stars
as candidate runaway stars (e.g.\ \cite{Hubrig2011a}).
Increasing the known number of magnetic objects with extreme rotation, which are probably 
products of a past binary interaction, is important to understand the magnetic field origin in massive stars.
The star $\zeta$\,Ophiuchi (=HD\,149757) of spectral 
type O9.5V is a well-known rapidly rotating runaway star, rotating almost at break-up
velocity with $v\,\sin\,i=400$\,km\,s$^{-1}$ \cite{Kambe}. 
The analysis of the FORS\,2 observations showed the presence of a weak magnetic field
with a reversal of polarity \cite{Hubrig2013} and an amplitude of about 100\,G. 
The resulting periodogram for the  magnetic  field  measurements  using all available lines showed
a dominating peak corresponding to a period of about  1.3\,d, which is 
roughly double the period of 0.643\,d determined by Pollmann \cite{Pollmann},
who studied the variation of the equivalent
width of the He~{\sc i}~6678 line. 

The presence of magnetic fields might change our whole picture about the evolution from O stars via WR stars to
supernovae or gamma-ray bursts. Neglecting magnetic fields could be one of the reasons why models
and observations of massive-star populations are still in conflict. Another potential importance of magnetic fields 
in massive stars concerns the dynamics of stellar winds.
A few years ago, Hubrig et al.\ (2016, {\sl submitted})  carried out FORS\,2 observations of a 
sample of Galactic WR stars including one WR star in the Large Magellanic Cloud.
Magnetic fields in WR stars are especially hard to detect because of 
wind-broadening of their spectral lines. Moreover, all photospheric lines are absent and 
the magnetic field is measured on emission lines formed
in the strong wind.
Remarkably, spectropolarimetric monitoring
of WR\,6, one of the brightest WR stars, revealed a sinusoidal nature of 
$\left<B_{\rm z}\right>$ variations with a period of 3.77\,d, with an amplitude of only 70--90\,G.

\section{Pulsating massive stars}\label{puls}

\begin{figure}
\centering
\includegraphics[width=0.40\textwidth]{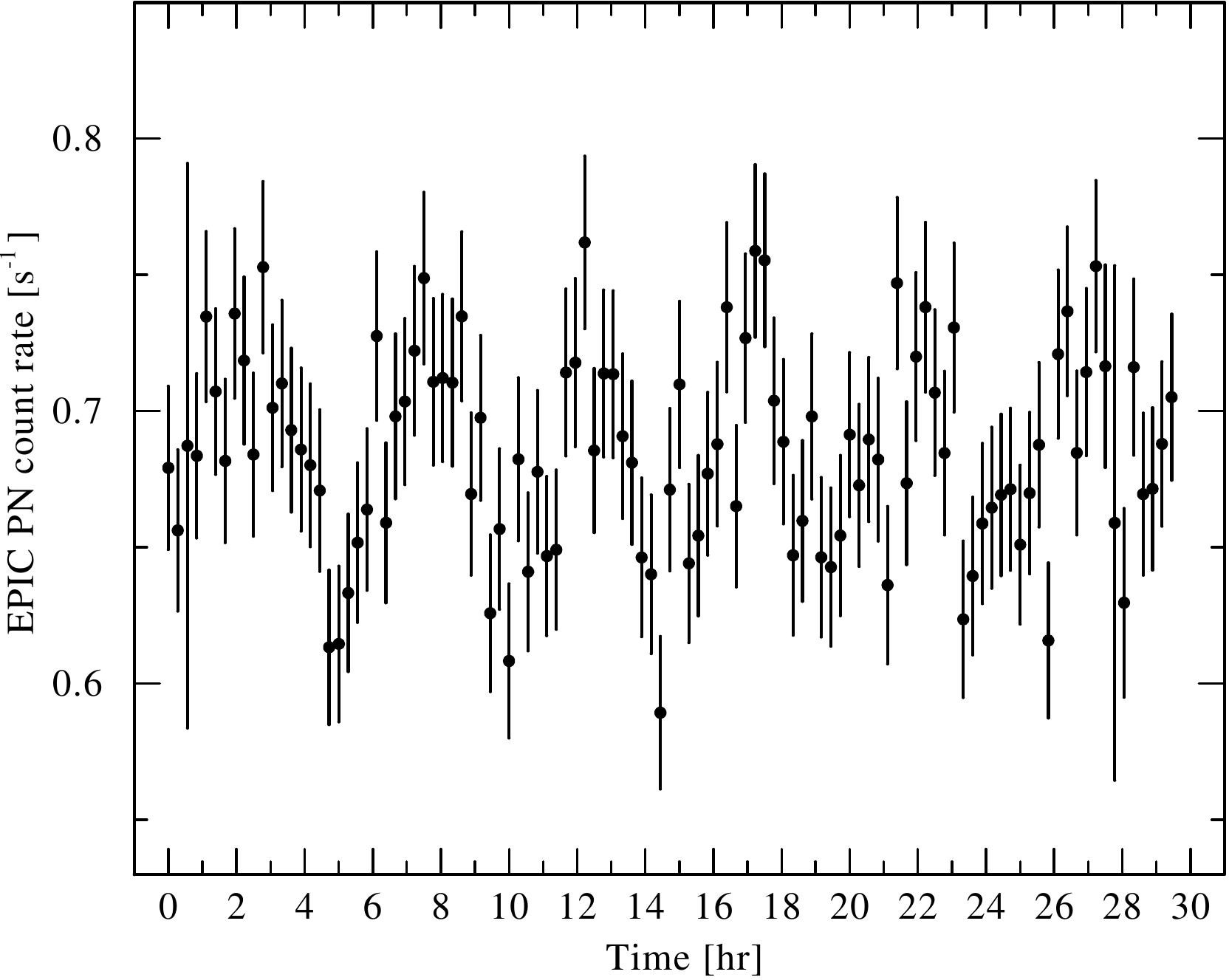}
\caption{
 X-ray light curve of $\xi^1$\,CMa in the 0.2\,keV -- 10.0\,keV 
(1.24\,\AA\,--\,62\,\AA) energy band, where the   background was
subtracted. The horizontal axis denotes  the time after the beginning of
the observation in hours. The data were binned to 1000\,s. 
The vertical axis shows the count rate as measured by the EPIC PN
camera. The  error bars (1$\sigma$) correspond to the combination 
of the error in the source counts and the background counts. 
}
\label{fig:lida1}
\end{figure}
\begin{figure}
\centering
\includegraphics[width=0.45\textwidth]{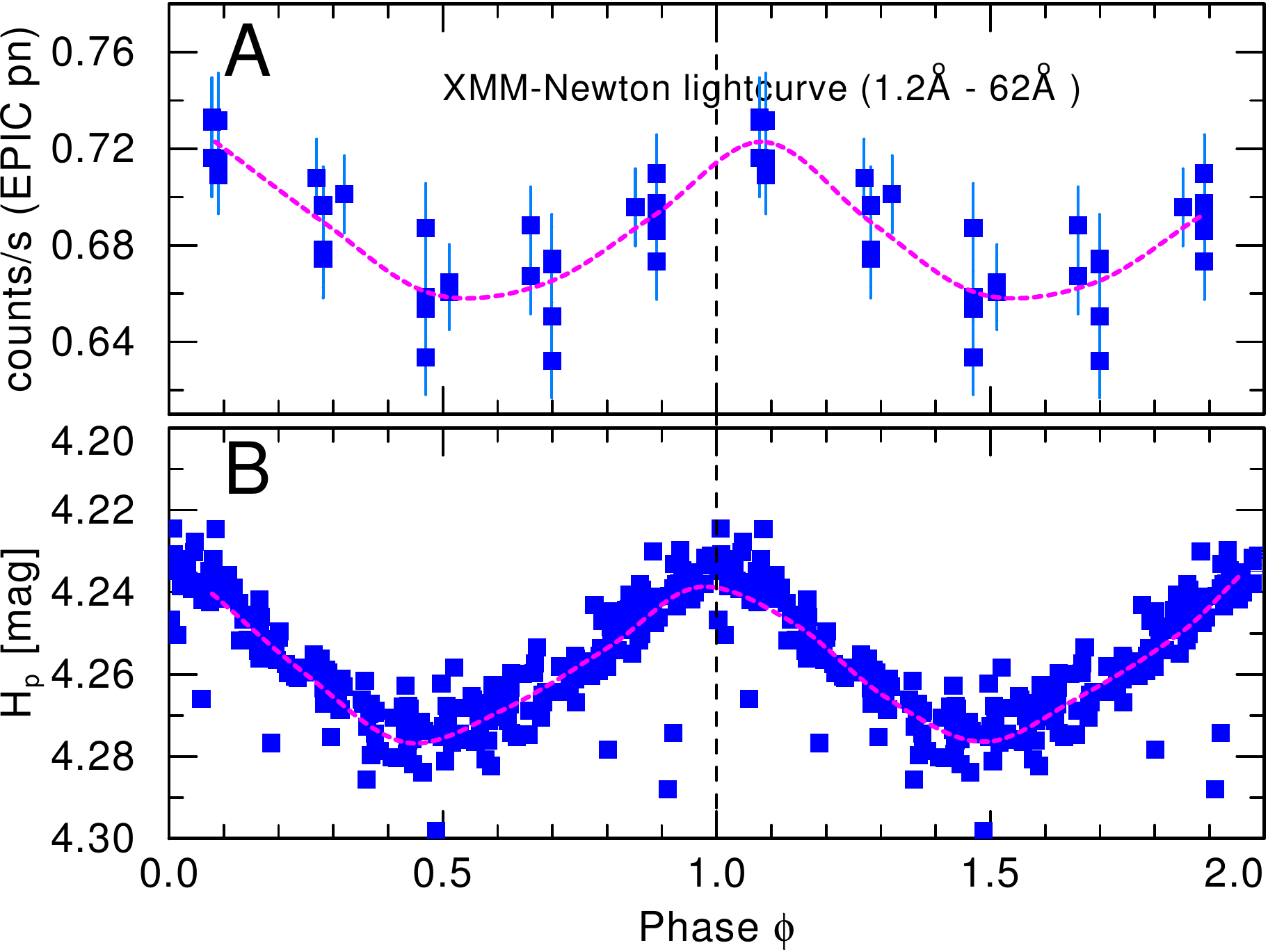}
\caption{
 X-ray (upper panel) and optical (lower panel)
light curves  of $\xi^1$\,CMa, phased with the stellar pulsation period. The
X-ray  light curve is produced from the data obtained with the  {\em
XMM-Newton} EPIC PN camera, using 1\,h binning.  The dashed red line interpolates  
the averages in phase bins of  $\Delta\phi=0.1$.  The lower panel shows the {\em Hipparcos}
Catalogue Epoch Photometry data. The abscissa is the magnitude $H_{\rm
p}$ in the {\em Hipparcos}  photometric system (330-900\,nm with maximum at
$\approx 420$\,nm).  The dashed red line interpolates  the averages.
}
\label{fig:lida2}
\end{figure}

Recent high-precision uninterrupted 
high-cadence space photometry using a number of satellites (e.g., WIRE,
MOST, CoRoT, Kepler, BRITE) led to a revolutionary change in the observational evaluation of variability
of massive stars.
Supported by results of photometric monitoring, it is expected that a large fraction of massive stars show
photometric variability due to either $\beta$\,Cep- or SPB-like pulsations, or stochastic $p$-modes,
or convectively-driven internal gravity waves.

High-resolution spectropolarimetric observations of pulsating stars frequently fail to show 
credible measurement results, if the whole sequence of subexposures at different retarder 
waveplate angles has a duration comparable to the timescale of the pulsation 
variability.
As an example, even for the bright fourth magnitude $\beta$~Cephei star $\xi^1$\,CMa with a pulsation
period of 5\,h, a full HARPS sequence of subexposures requires about 30\,min.
In contrast, one FORS\,2 observation of the same star lasts less than 10\,min.
Owing to the strong changes in the line profile positions and the shapes in the spectra of pulsating stars, a method using 
spectra averaged over all subexposures leads to erroneous wavelength shifts and thus to wrong values for the 
longitudinal magnetic field.

For the first time, FORS\,1 magnetic field surveys of 
slowly pulsating B (SPB) stars  and $\beta$\,Cephei  stars  
were  carried out from 2003 to 2008. As a number of pulsating stars showed the presence of a magnetic field, 
our observations implied that $\beta$~Cephei and SPB stars can no 
longer be considered as classes of non-magnetic pulsators.
Notably, although the presence of magnetic fields in these stars is already known for more than ten years, 
the effect of these fields on the oscillation properties is not yet understood and remains to be studied.
$\xi^1$\,CMa, discovered as magnetic with FORS\,1 observations long ago,
is still the record holder with the strongest mean longitudinal magnetic field among the $\beta$~Cephei stars
of the order of 300--400\,G \cite{Hubrig2006}.
Using FORS\,2 measurements obtained in service mode in 2009/10,
Hubrig et al.\ \cite{Hubrig2011b} detected a rotational modulation 
of its magnetic field with a period of about 2.19\,d and estimated a magnetic dipole strength of about 5.3\,kG.

Fully unexpected, observations of this particular star with the {\em XMM-Newton} telescope revealed for the first time 
X-ray pulsations with the same period as the stellar radial pulsation \cite{Oskinova}.
In Figs.~\ref{fig:lida1} and  \ref{fig:lida2},
we present the observed X-ray light curve and the X-ray/optical light curves
phased with the pulsation period.
This first discovery of X-ray pulsations from a non-degenerate, massive
star, stimulates theoretical considerations for the physical
processes  operating in magnetized stellar winds.

Observations of pulsating stars also allowed the first detection of a magnetic field in another  $\beta$~Cephei star, 
$\epsilon$~Lup \cite{Hubrig2009}, which is an SB2 system and recently received attention
due to the presence of a magnetic field in both components.
Since binary systems with magnetic components are rather rare, the detection
of a magnetic field in this system using low resolution FORS spectropolarimetry indicates
the potential of FORS\,2 also for magnetic field searches in binary or multiple systems. 

\section{Improvements in the measurement techniques}

During the last years, the measurement strategy for high-resolution and low-resolution
spectropolarimetric observations was modified in many aspects. To measure the mean longitudinal
magnetic fields in high-resolution polarimetric spectra obtained with ESPaDOnS, NARVAL, and HARPS
most teams are using the moment technique introduced by Mathys \cite{mathys} and the Least-Squares Deconvolution (LSD)
introduced by Donati et al.\ \cite{donati}.
In the last years, Carroll et al.\ \cite{carroll2012} developed 
the  multi-line Singular Value Decomposition (SVD) method for Stokes Profile Reconstruction.
The basic idea of SVD is similar to the Principal Component Analysis  approach, where 
the similarity of the individual Stokes $V$ profiles allows one to describe the most coherent and 
systematic features present in all spectral line profiles as a projection onto a small number of 
eigenprofiles (e.g.\ \cite{carroll2009}). The excellent potential of the SVD
method, especially in the analysis of extremely weak fields, e.g.\ in the Herbig Ae/Be star PDS\,2,
was recently demonstrated by Hubrig et al.\ (\cite{Hubrig2015b}; right side of their Fig.\,4).
  
In the reduction process of low-resolution spectropolarimetric observations, 
Hubrig et al.\ \cite{Hubrig2014} perform  rectification of the $V/I$ spectra and calculate 
null profiles, $N$, as pairwise differences from all available 
$V$ profiles.  From these, 3$\sigma$-outliers are identified and used to clip 
the $V$ profiles. This removes spurious signals, which mostly come from cosmic
rays, and also reduces the noise. A full description of the updated data 
reduction and analysis will be presented in a paper by Sch\"oller et 
al.\ ({\sl in preparation}). 

The mean longitudinal magnetic field, $\left<B_{\rm z}\right>$, is defined by the slope of the 
weighted linear regression line through the measured data points, where
the weight of each data point is given by the squared signal-to-noise ratio
of the Stokes $V$ spectrum. The formal $1\sigma$ error of 
$\left<B_{\rm z}\right>$ is obtained from the standard relations for weighted 
linear regression. This error is inversely proportional to the rms  
signal-to-noise ratio of Stokes $V$.
Finally, we apply the factor
$\sqrt{\chi^2_{\rm min}/\nu}$ to the error determined from the 
linear regression, if larger than 1.

Since 2014, Hubrig et al.\ \cite{Hubrig2014} also implement
the Monte-Carlo bootstrapping technique,
where they typically generate $M=250\,000$ statistical variations of the 
original dataset, and analyze the resulting 
distribution $P(\left<B_{\rm z}\right>)$ of the $M$ regression results. Mean 
and standard deviation of this distribution are identified with the most 
likely mean longitudinal magnetic field and its $1\sigma$ error, 
respectively. The main advantage of this method is that it provides an 
independent error estimate. 

A number of discrepancies in the published measurement accuracies has been reported by Bagnulo et al.\ \cite{bag},
who used the ESO FORS\,1 pipeline to reduce the full content of the FORS\,1 archive.
The same authors already published a few similar papers in the last years suggesting that
very small instrument flexures, negligible in most of the instrument applications, may be responsible for some 
spurious magnetic field detections, and that FORS detections may be considered reliable only at a level 
greater than 5$\sigma$.
However, no report on the presence of flexures from any astronomer 
observing with the FORSes was ever published in the past.
The authors also discuss the impact of seeing, if the exposure time
is comparable with the atmospheric coherence time, which they incorrectly assume to be in seconds and
not in milliseconds.
In the most recent  work do the authors present for the first time the level of 
intensity fluxes for each image
and report which spectral regions were used for the magnetic field measurements. However, no fluxes for 
left-hand and right-hand polarized spectra are available, thus the reproduction of their measurements is not 
possible.
Notably, already small changes in the spectral regions selected for the measurements can 
have a significant impact on the measurement results \cite{Hubrig2004}.

\begin{figure}
\centering
\includegraphics[angle=0,width=0.23\textwidth]{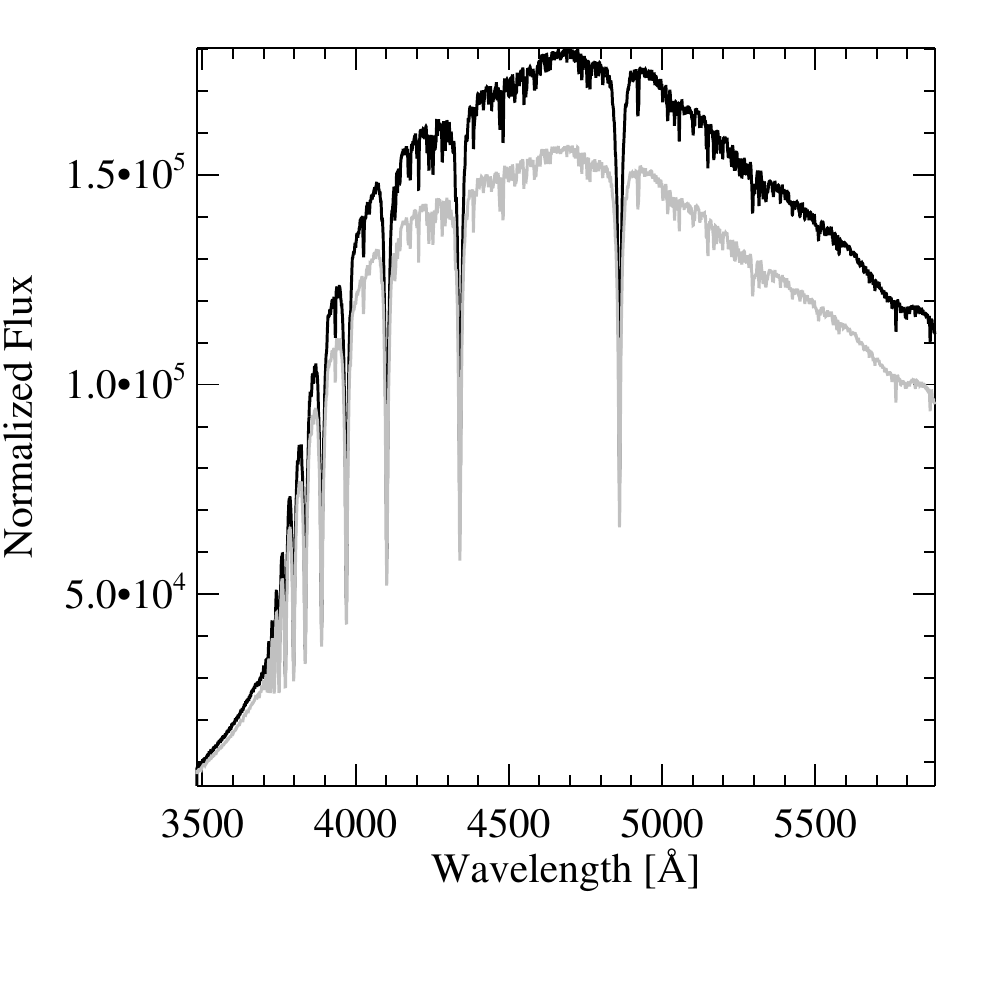}
\includegraphics[angle=0,width=0.23\textwidth]{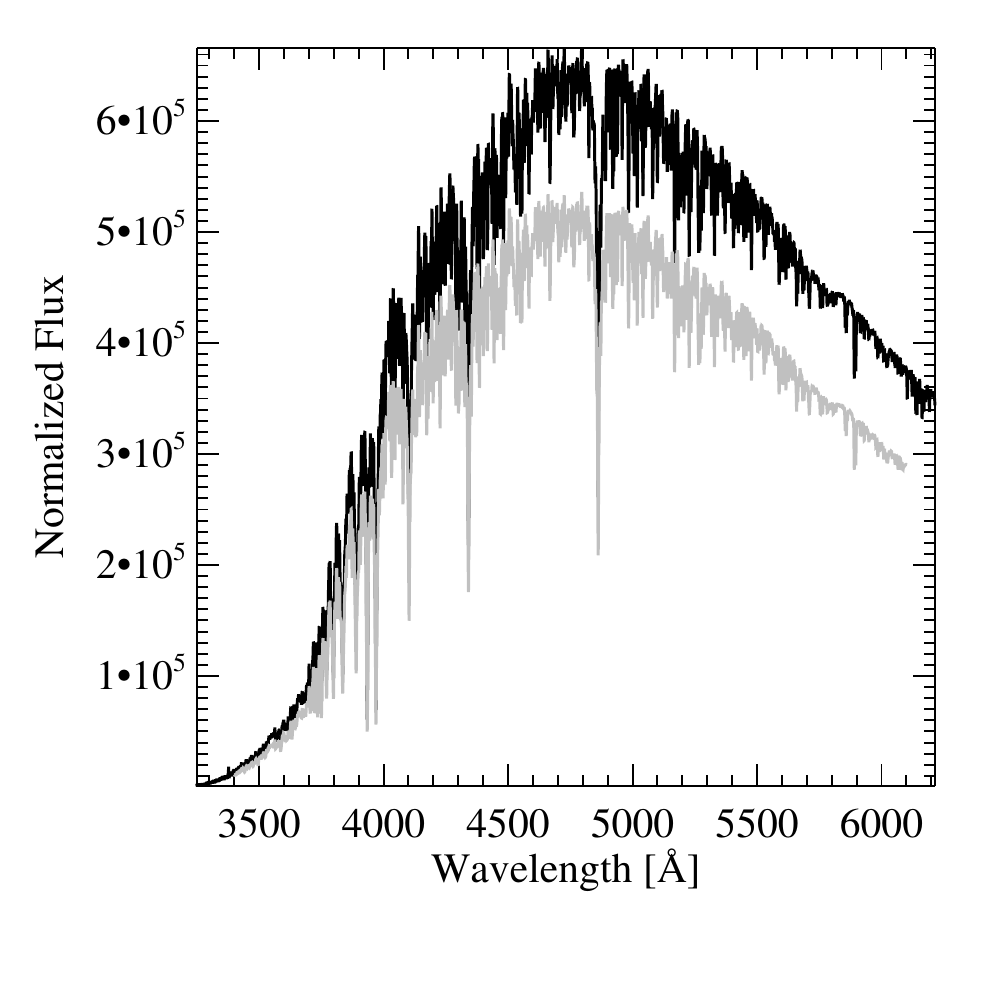}
\includegraphics[angle=0,width=0.23\textwidth]{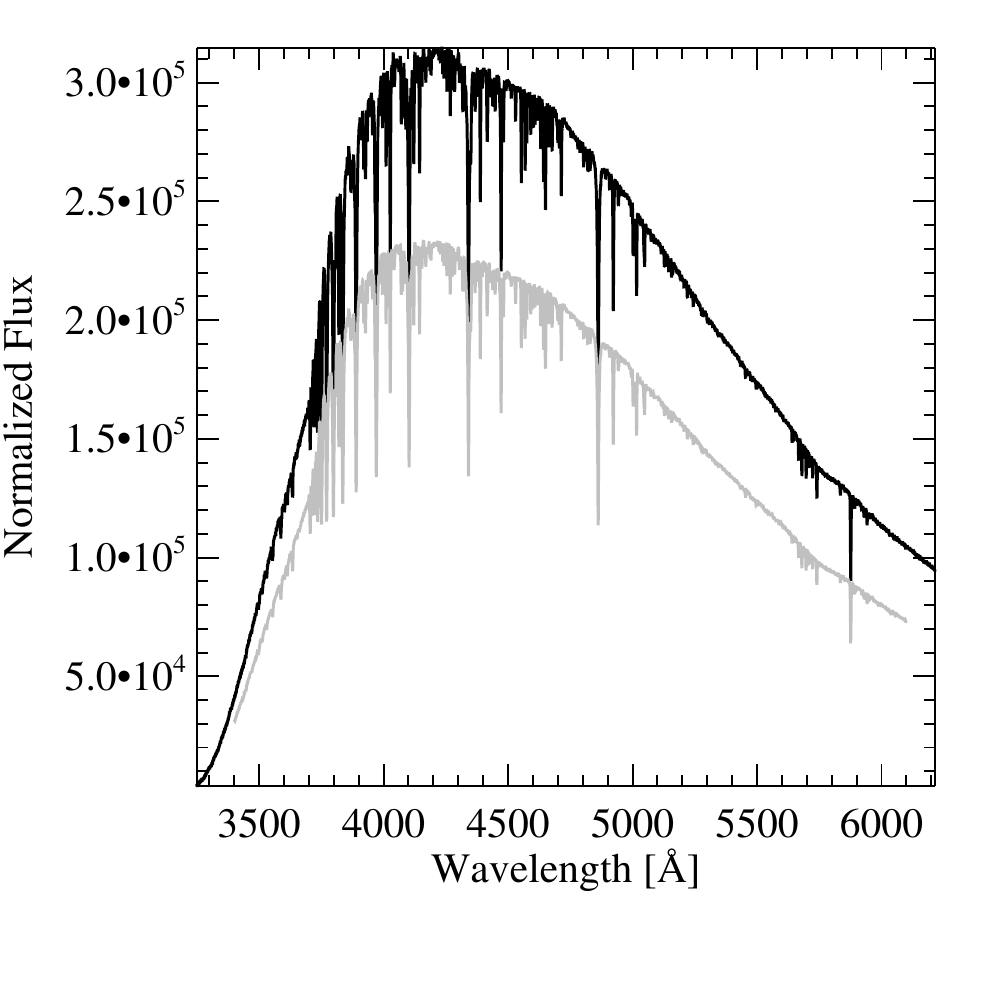}
\includegraphics[angle=0,width=0.23\textwidth]{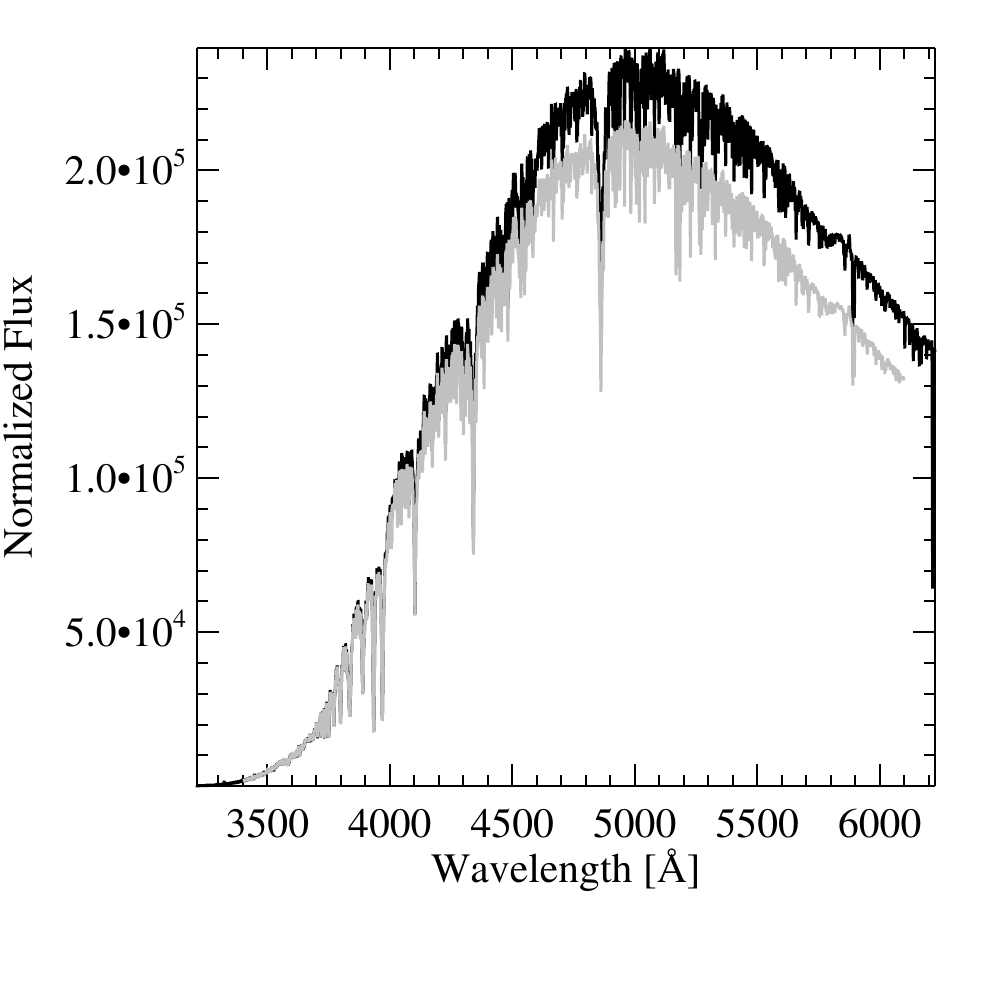}
\caption{
Fluxes extracted by Bagnulo et al.\ \cite{bag} (grey color) compared to those using our own pipeline (black color).
The differences in the fluxes are presented (from left to right) for the HgMn star $\alpha$~And, 
the $\delta$~Scuti star HD\,21190, the nitrogen rich early B-type star HD\,52089, and the Herbig Ae star PDS\,2.
All these stars were announced in studies by Hubrig et al.\ as magnetic.
}
\label{fig:fluxes}
\end{figure}

Since the measurement accuracies predominantly depend on 
photon noise, an improper extraction of the spectra, for instance the use of smaller extraction
windows, would explain why Bagnulo et al.\ \cite{bag} disregarded 3$\sigma$
detections by other authors.
Indeed, the inspection of the levels of intensity fluxes for each subexposure compiled in the catalog 
of Bagnulo et al.\ \cite{bag} shows that their levels are frequently lower, down to 70\% in comparison to
those obtained in our studies.
In Fig.~\ref{fig:fluxes}, we present the comparison of fluxes for a few stars for 
which detections were achieved and published by Hubrig et al.\ during the last years.
It is obvious that the detection of 
weak magnetic fields is especially affected if the extracted fluxes are low.
From the consideration of the SNR values presented by Bagnulo et al.\ \cite{bag}, we also noted
that emission lines are not taken into account during the measurements. The reason for 
this is not clear to us, as there is no need to differentiate between 
absorption and emission lines: the used relation between the Stokes $V$ 
signal and the slope of the spectral line wing 
holds for both type of lines, so that the signals 
of emission and absorption lines add up rather than cancel.

\section{Summary}\label{outlook}

To increase the reliability of magnetic field detections, but also 
to carry out a quantitative atmospheric analysis and to probe spectral variability, it is 
certainly helpful to follow up FORS\,2 detections with high-resolution HARPS observations. 
To our knowledge, the only collaboration that uses  FORS\,2 and HARPS to monitor 
magnetic fields is the BOB (``B-fields in OB stars'') collaboration \cite{Morel2014},
which is focused on the search of magnetic fields in massive stars.
Combining observations with different instruments
allowed the BOB collaboration to report during the last couple of years the presence of magnetic fields
in a number of massive stars.
As an example, the first detection of a magnetic field in the single slowly rotating O9.7\,V star HD\,54879
was achieved with FORS\,2 and follow-up HARPS observations could show that HD\,54879  is,  so  far,  
the  strongest  magnetic  single  O-type star  detected  with  a  stable  and  normal  optical  spectrum
\cite{Castro}.


\begin{thebibliography}{9}



%
%
%
%
%
%
%
%

\bibitem{Naze}
\textit{Y.\,Naz\'e, A.\,Ud-Doula, M.\,Spano, et al.}, A\&A, \textbf{520}, A59, 2010.

\bibitem{Wade}
\textit{G.A.\,Wade, G.\,Grunhut, G.\,Gr\"afener, et al.}, MNRAS, \textbf{419}, 2459, 2012.

\bibitem{Hubrig2008}
\textit{S.\,Hubrig, M.\,Sch{\"o}ller, R.S.\,Schnerr, et al.}, A\&A, \textbf{490}, 793, 2008.

\bibitem{Hubrig2013}
\textit{S.\,Hubrig, M\,Sch{\"o}ller, I.\,Ilyin, et al.}, A\&A, \textbf{551}, A33, 2013.

\bibitem{Walborn}
\textit{N.R.\,Walborn}, AJ, \textbf{78}, 1067, 1973.

\bibitem{Hubrig2015a}
\textit{S.\,Hubrig, M.\,Sch{\"o}ller, A.F.\,Kholtygin, et al.}, MNRAS, \textbf{447}, 1885, 2015.

\bibitem{rivinius}
\textit{T.\,Rivinius, R.H.D.\,Townsend, O.\,Kochukhov, et al.}, MNRAS, \textbf{429}, 177, 2013. 

\bibitem{Hubrig2011a}
\textit{S.\,Hubrig, N.V.\,Kharchenko, M.\,Sch{\"o}ller}, Astr.\ Nachr., \textbf{332}, 65, 2011.

\bibitem{Kambe}
\textit{E.\,Kambe, R.\,Hirata, H.\,Ando},  A\&A, \textbf{273}, 435, 1993.

\bibitem{Pollmann}
\textit{E.\,Pollmann}, IBVS, \textbf{6034}, 1, 2012.

\bibitem{Hubrig2006}
\textit{S.\,Hubrig, M.\,Briquet, M.\,Sch{\"o}ller, et al.}, MNRAS, \textbf{369}, L61, 2006.

\bibitem{Hubrig2011b}
\textit{S.\,Hubrig, I.\,Ilyin, M.\,Sch\"oller, et al.}, ApJ,  \textbf{726}, L5, 2011.

\bibitem{Oskinova}
\textit{L.M.\,Oskinova, Y.\,Naz\'e, H.\,Todt, et al.},  Nature Commun., \textbf{5}, 4024, 2014.

\bibitem{Hubrig2009}
\textit{S.\,Hubrig, M.\,Briquet, P.\,De Cat, et al.},  Astr.\ Nachr., \textbf{330}, 317, 2009.

\bibitem{mathys}
\textit{G.\,Mathys}, IAUC, \textbf{138}, 232, 1993.

\bibitem{donati}
\textit{J.F.\,Donati, M.\,Semel, B.D.\,Carter, et al.}, MNRAS, \textbf{291}, 658, 1997.

\bibitem{carroll2012}
\textit{T.A.\,Carroll, K.G.\,Strassmeier, J.B.\,Rice, A.\,K\"unstler}, A\&A, \textbf{548}, A95, 2012.

\bibitem{carroll2009}
\textit{T.A.\,Carroll, M.\,Kopf, K.G.\,Strassmeier, I.\,Ilyin}, IAUS, \textbf{259}, 633, 2009.

\bibitem{Hubrig2015b}
\textit{S.\,Hubrig, T.A.\,Carroll, M.\,Sch\"oller, I.\,Ilyin}, MNRAS, \textbf{449}, L118, 2015.

\bibitem{Hubrig2014}
\textit{S.\,Hubrig,  M.\,Sch\"oller, A.F.\,Kholtygin}, MNRAS,  \textbf{440}, 1779, 2014.

\bibitem{bag}
\textit{S.\,Bagnulo, L.\,Fossati, J.D.\,Landstreet, C.\,Izzo},  A\&A, \textbf{583}, A115, 2015.

\bibitem{Hubrig2004}
\textit{S.\,Hubrig, D.W.\,Kurtz, S.\,Bagnulo, et al.}, A\&A,  \textbf{415}, 661, 2004.

\bibitem{Morel2014}
\textit{
Morel, T., Castro, N., Fossati, L., et al.}, The Messenger,  \textbf{157}, 27, 2014.

\bibitem{Castro}
\textit{N.\,Castro, L.\,Fossati, S.\,Hubrig, et al.}, A\&A, \textbf{581}, A81, 2015.

\end{thebibliography}
\end{document}